\documentclass[11pt,a4paper]{article}
\pdfoutput=1

\usepackage{amsmath}
\usepackage{amssymb}
\usepackage{cite}
\usepackage{graphicx}
\usepackage{bbm}
\usepackage{color}
\usepackage{url}

\newcommand{\be}{\begin{equation}}  
\newcommand{\ee}{\end{equation}}  
\newcommand{\ol}[1]{\overline{#1}}
\newcommand{\hc}{+\,\mathrm{h.c.}}
\newcommand{\vev}[1]{\langle #1 \rangle}

\newcommand{\SU}[1]{\ensuremath{\mathrm{SU}(#1)}}
\newcommand{\U}[1]{\ensuremath{\mathrm{U}(#1)}}
\newcommand{\into}{\ensuremath{\,\rightarrow\,}}
\newcommand{\tr}{\operatorname{tr}}

\newcommand{\wt}[1]{\widetilde{#1}}

\newcommand{\MP}{\ensuremath{M_\mathrm{P}}}

\renewcommand{\d}{\ensuremath{\mathrm{d}}}

\title{
\textbf{Coscattering in next-to-minimal dark matter and split supersymmetry} 
\vspace{2ex}
}

\author{\large F.~Br\"ummer\\[1ex]
\textit{\normalsize LUPM, UMR5299, Universit\'e de Montpellier and CNRS}\\
\textit{\normalsize 34095 Montpellier, France}\\[1ex]
\vspace{3ex}
}
\date{}
\begin{document}

\maketitle

\begin{abstract} \noindent
In some models of thermal relic dark matter, the relic abundance may be set by inelastic scattering processes (rather than annihilations) becoming inefficient as the universe cools down. This effect has been called coscattering. We present a procedure to numerically solve the full momentum-dependent Boltzmann equations in coscattering, which allows for a precise calculation of the dark matter relic density including the effects of early kinetic decoupling. We apply our method to a simple model, containing a fermionic SU(2) triplet and a fermionic singlet with electroweak-scale masses, at small triplet-singlet mixing. The relic density can be set by either coannihilation or, at values of the mixing angle $\theta\lesssim 10^{-5}$, by coscattering. We identify the parameter ranges which give rise to the observed relic abundance. As a special case, we study bino-like dark matter in split supersymmetry at large $\mu$.
\end{abstract}

\section{Introduction}

Weakly interacting massive particles (WIMPs) with electroweak-scale masses are among the best-motivated candidates for particle dark matter. In WIMP models, the dark matter particle is in equilibrium with the thermal bath of Standard Model particles at early cosmic times, until the rates of WIMP-number changing processes drop below the Hubble expansion rate. Below the temperature of this so-called freeze-out, the dark matter number density becomes effectively constant.

It has long been known that the presence of additional states $\psi$ whose masses are close to the dark matter mass can significantly affect the prediction for the dark matter relic density. This is the case when these states are able to coannihilate with the dark matter particle $\chi$  \cite{Griest:1990kh}. It is also well known, but perhaps less universally appreciated, that coannhilations can lead to the observed relic density even when the actual $\chi\chi\into X$ and $\chi\psi\into X$ annihilation cross-sections are several orders of magnitude below the typical electroweak cross-section. All that is needed is for the coannihilation partners $\psi$ to annihilate efficiently among themselves, and for the dark matter particle to remain in equilibrium via inelastic scattering $\chi X\into \psi X'$, such that any $\chi$ overdensity is rapidly converted into a $\psi$ overdensity which is subsequently washed out.

It is interesting to study scenarios where such inelastic scattering processes start becoming inefficient before $\psi\psi\into X$ annihilations does \cite{DAgnolo:2017dbv, Garny:2017rxs}. In that case, the usual coannihilation formalism \cite{Griest:1990kh, Gondolo:1990dk, Edsjo:1997bg} fails, since one of its core assumptions is that $\chi$ is always in equilibrium with $\psi$. Having the dark matter relic abundance dictated by the freeze-out of inelastic scattering processes rather than of coannihilations has been dubbed ``coscattering'' in \cite{DAgnolo:2017dbv}. This mechanism has since been explored in the context of several different models \cite{Garny:2018icg, Cheng:2018vaj, Junius:2019dci, Kim:2019udq}.

The conditions for coscattering to occur are thus that (1) the dark matter particle $\chi$ is in equilibrium with the thermal bath at early times, (2) there exists one or more state(s) $\psi$ carrying the same charge as $\chi$ under the symmetry which stabilizes $\chi$, (3) the couplings and masses are such that $\chi\psi\into X$ annihilations as well as $\chi\chi\into X$ annihilations decouple earlier than $\chi X\into\psi X'$ conversions, and these in turn decouple earlier than $\psi\psi\into X$ annihilations.\footnote{Decays and inverse decays may or may not be active \cite{Garny:2017rxs}.} It has been argued \cite{DAgnolo:2017dbv} that, in the case that $\psi$ and $\chi$ are coupled via an extra mediator particle $\phi$ with $m_\phi+m_\psi>2\,m_\chi$, the dark matter mass should be exponentially below the weak scale. While this could open up interesting directions for model building, in the present work our example models are closer to standard WIMP models: $m_\chi$ and $m_\psi$ are both of the order of a few 100 GeV, and no mediators are present other than the Standard Model particles.

More specifically, we will study coscattering in one of the simplest models where it can occur, namely, the singlet-triplet next-to-minimal dark matter model of \cite{Bharucha:2017ltz,Bharucha:2018pfu}. A special case of this model is split supersymmetry \cite{ArkaniHamed:2004fb, Giudice:2004tc} with a bino-like lightest neutralino and a large higgsino mass parameter. Besides the Standard Model particles, the model contains only two multiplets with electroweak-scale masses, namely, a fermionic singlet $\chi$ and a fermionic $\SU{2}$ triplet $\psi$. We will give  Majorana masses of the order of the electroweak scale to both these fields, with $\chi$ being slightly lighter than $\psi$, and impose a $\mathbb{Z}_2$ symmetry under which they are odd. This latter symmetry forbids any renormalizable interactions between $\chi$ and the Standard Model states, but allows for the dimension-5 operator $\frac{1}{\Lambda}\chi\psi H^\dag H$ which mixes the neutral components of $\chi$ and $\psi$ (here $H$ is the Standard Model Higgs doublet and $\Lambda$ is a cutoff scale). At large $\Lambda$, or equivalently small mixing angles, $\chi\chi\into X$ annihilations as well as $\chi\psi\into X$ coannihilations quickly become negligible as the temperature drops below the dark matter mass. However, $\psi\psi\into X$ annihilations as well as $\chi X\into\psi X'$ scattering remain efficient at first, the former because they are not suppressed by the mixing angle and the latter because of its less severe Boltzmann suppression (taking $X$ and $X'$ to be relativistic Standard Model particles). Depending on the masses and the mixing angle, either $\psi\psi$ annihilation or inelastic scattering may be the first process to start decoupling as the temperature drops further, thus giving rise to either coannihilation or coscattering.

A subtlety of the coscattering phase is early kinetic decoupling, as local (or ``kinetic'') equilibrium is lost through the decoupling of the very same inelastic scattering processes which determine the dark matter relic abundance. Therefore, another key assumption of the usual coannihilation formalism is violated in coscattering scenarios, since the momentum distribution of dark matter at freeze-out is not necessarily an equilibrium (Maxwell-Boltzmann) distribution. This needs to be taken into account properly when predicting the thermal relic density, as emphasized already in \cite{DAgnolo:2017dbv, Garny:2017rxs}, and may significantly change the result in certain cases \cite{Duch:2017nbe, Binder:2017rgn}.

The aim of this paper is to establish a framework allowing for an accurate computation of the dark matter relic abundance in coscattering models, including the effects of early kinetic decoupling, building on the formalism proposed in \cite{Garny:2017rxs}. We will apply our method to analyze the singlet-triplet model at and beyond the transition between coannihilation and coscattering. 

After briefly recalling the essential properties of singlet-triplet next-to-minimal dark matter, we will proceed to review and to further develop the formalism for calculating the dark matter relic density. We will then present some numerical results in the singlet-triplet model, and as a special case, we will discuss the parameter space for bino-like dark matter in split supersymmetry. We will conclude with some remarks about present constraints and possible future experimental signatures relevant to these models, and with a brief summary.

\section{Singlet-triplet next-to-minimal dark matter}
\label{sec:tsnmdm}

We will proceed with a brief review of the singlet-triplet model studied in \cite{Bharucha:2017ltz, Bharucha:2018pfu}, emphasizing its possible origins in split supersymmetry. We add to the Standard Model a fermionic singlet $\chi$ as well as its fermionic coannihilation partner $\psi$ with quantum numbers $({\bf 1}, {\bf 3})_{0}$ under $\SU{3}\times\SU{2}\times \U{1}$. The most general Lagrangian compatible with this particle content and a $\mathbb{Z}_2$ symmetry under which $\chi$ and $\psi$ are odd is
\be
{\cal L}={\cal L}_{\rm SM} + i\chi^\dag\bar\sigma^\mu\partial_\mu\chi+i\psi^\dag\bar\sigma^\mu D_\mu\psi+\frac{1}{2}\left(m\chi\chi + M\psi\psi\hc\right)+{\cal L}_{5}+{\cal L}_{\geq 6}
\ee
where ${\cal L}_{5}$ contains the dimension-5 operators
\be\label{eq:Lnonren}
{\cal L}_{5}=\frac{1}{2}\frac{\kappa}{\Lambda}\psi\psi H^\dag H +\frac{1}{2}\frac{\kappa'}{\Lambda}\chi\chi H^\dag H +\frac{\lambda}{\Lambda}\chi\psi^a H^\dag\tau^a H\hc+\ldots
\ee
Here $H$ is the Standard Model Higgs doublet. Apart from the Standard Model Weinberg operator these are the only possible dimension-5 terms. We will assume that dimension-6 and higher operators are negligible in the following. We will restrict our study to real parameters and choose $M>0$ without loss of generality. Since the dark matter particle is to be $\chi$-like, we assume that $M>|m|$.

This model can be UV-completed with a $\mathbb{Z}_2$-odd Dirac fermion doublet $\Psi$ with hypercharge $\frac{1}{2}$ whose mass is of the order of $\Lambda$. Identifying $\chi\sim\wt B$, $\psi\sim\wt W$, $\Psi\sim\left(\wt H_u^\dag, \wt H_d\right)$, the particle content is then exactly the low-energy particle content of split supersymmetry, except for a gluino which will play no role in our considerations (assuming that it is heavy enough not to conflict with collider bounds).\footnote{Here, as usual in the supersymmetric literature, $\wt B$, $\wt W$ and $\wt H_{u,d}$ denote the fermionic superpartners of the $B$, $W$ and Higgs bosons respectively.} In split supersymmetry, the cutoff scale $\Lambda$ is identified with the Higgsino mass scale $|\mu|$, the singlet and triplet masses $m$ and $M$ with the supersymmetry-breaking gaugino mass parameters $M_1$ and $M_2$, and the Wilson coefficients $\lambda$, $\kappa$ and $\kappa'$ in Eq.~\eqref{eq:Lnonren} are given at the renormalization scale $|\mu|$ by the matching conditions
\be\label{eq:ssmatching}
\lambda=\left(\tilde g_u\tilde g_d'+\tilde g_d\tilde g_u'\right){\rm sign}\,\mu\,,\qquad\kappa=\tilde g_u \tilde g_d\,{\rm sign}\,\mu\,,\qquad \kappa'=\tilde g_u' \tilde g_d'\,{\rm sign}\,\mu\,.
\ee
Here $\tilde g_{u,d}$ and $\tilde g'_{u,d}$ are the usual gaugino-Higgsino-Higgs Yukawa couplings of split supersymmetry \cite{Giudice:2004tc}. It is understood that the split supersymmetry model will be embedded in a fully supersymmetric model at an even higher scale $M_{\rm SUSY}$ which is the mass scale of the remaining superpartners. Hence we are assuming the mass hierarchy
\be
M_{\rm SUSY}\gg|\mu|\gg M_2 > |M_1|\,.
\ee
Since we are not assuming any particular model of supersymmetry breaking, this is a possible choice of parameters. In fact, the origins of the $\mu$ parameter being supersymmetric, there is no a priori reason for it to be correlated with any of the supersymmetry-breaking parameters (up to the issue of radiative corrections, which we will comment on in Section \ref{sec:split}).

Regardless of the nature of the UV completion, after electroweak symmetry breaking and upon replacing the Higgs field by its vacuum expectation value
\be
\vev H=\left(\begin{array}{c} 0 \\ v \end{array}\right)\,,\qquad v=174\,{\rm GeV}\,,
\ee
the first two terms in Eq.~\eqref{eq:Lnonren} will induce a shift in the effective $\chi$ and $\psi$ mass parameters respectively, which we will absorb in the definitions of $m$ and $M$. The third term will cause the neutral component of $\psi$ to mix with $\chi$. The mixing angle is
\be\label{eq:mixing}
\theta=\frac{|\lambda|\, v^2}{2\Lambda (M-m)}\,.
\ee
We will study $\lesssim {\cal O}(1)$ Wilson coefficients $\lambda$ and mass differences $M-m$ which are not parametrically smaller than the electroweak scale, such that the mixing angle is roughly of the order $v/\Lambda$.

The mass eigenstates in the dark matter sector are then a neutral Majorana fermion $\chi^0$ which is mostly $\chi$-like, a neutral Majorana fermion $\psi^0$ which is mostly $\psi$-like, and a purely $\psi$-like charged Dirac fermion $\psi^\pm$. (In split supersymmetry, these would correspond to a bino-like lightest neutralino $\chi^0_1$, a wino-like next-to-lightest neutralino $\chi^0_2$ and a wino-like chargino $\chi^\pm_1$ respectively, with the higgsino-like states at the higher mass scale $|\mu|$). In the limit of sending the cutoff scale to infinity, $\chi^0$ becomes purely $\chi$-like and completely inert. The mass degeneracy between $\psi^0$ and $\psi^\pm$ is lifted by $\psi^0-\chi^0$ mixing and by electroweak loops, the latter inducing a mass difference of around $160$ MeV between the neutral and the charged $\psi$-like states \cite{Ibe:2012sx}. The mixing-induced mass difference is even smaller for the mixing angles of interest to us.

\section{Coannihilation and coscattering in the singlet-triplet model}
\label{sec:coco}

The fermionic singlet-triplet model described in Section \ref{sec:tsnmdm} is a prime example of a model which can produce the dark matter relic density via either coannihilation or coscattering. While coannihilation in the singlet-triplet (or wino-bino) model has been extensively discussed in the literature (see e.g.~\cite{Baer:2005jq, ArkaniHamed:2006mb, Ibe:2013pua, Harigaya:2014dwa, Bharucha:2017ltz, Bramante:2015una, Yanagida:2019evh}), the coscattering phase has not been studied previously. 

We take $M$ and $m$ of the order of a few 100 GeV and degenerate within $<10\%$. The effective dark matter number density will be depleted by  $\chi\chi\into X$, $\chi\psi\into X$ and $\psi\psi\into X$ annihilation processes (where $\psi$ is any of the triplet-like states $\psi^0, \psi^+$ or $\psi^-$, and here and in the following we write $\chi$ instead of $\chi^0$ to simplify notation --- it should be understood that we are dealing with a mass eigenstate which contains a small admixture of $\psi^0$). As the cutoff scale $\Lambda$ becomes large and the mixing angle becomes small, $\theta\ll 1$, the annihilation cross-sections involving $\chi$ become subdominant, while $\psi\psi\into X$ annihilations are independent of $\Lambda$. The dark matter particle $\chi$ is kept in chemical equilibrium as long as $\chi X\into\psi X'$ scattering remains efficient; these processes are also suppressed by $\Lambda$ but, in contrast to annihilations, they are only singly-Boltzmann suppressed if $X$ and $X'$ are relativistic Standard Model states. 

In the coannihilation phase of the model, the mixing angle is large enough to keep $\chi$ in equilibrium until $\psi\psi\into X$ annihilations freeze out. For this phase, the resulting dark matter relic density can be reliably computed using standard public codes. By contrast, in the coscattering phase, the mixing angle is so small that the $\chi X\into\psi X'$ scattering rate drops below the Hubble rate before the $\psi\psi\into X$ annihilation rate does. The $\chi$ number density will then decrease much more slowly and eventually become constant, while the $\psi$-like particles will continue to annihilate until the freeze-out temperature $T_f\approx M/25$ is reached. At even lower temperatures, the few remaining $\psi$-like particles will ultimately slowly convert or decay into $\chi$, thus increasing the final dark matter relic abundance by a small amount. 

We proceed by presenting the details of the calculation of the thermal relic abundance, reviewing and extending the formalism developed in \cite{Garny:2017rxs} as we go along. Our starting point is the Boltzmann equation in a Friedmann-Robertson-Walker universe,
\be\label{eq:fullBoltzmann}
\left(\partial_t-H\,{\bf p}_\chi\cdot\nabla_{{\bf p}_\chi}\right)f_\chi ({\bf p}_\chi, t)=\frac{1}{E_\chi} C[f_\chi]\,.
\ee
Here $t$ is cosmic time and $H$ is the Hubble parameter, which, in a radiation-dominated universe, is given as a function of temperature by 
\be
H=\frac{T^2}{\MP}\pi \sqrt{\frac{g_*}{90}}
\ee
with $g_*(T)$ the effective number of degrees of freedom and $M_P=2.435\times 10^{18}$ GeV the reduced Planck mass. Moreover, $f_\chi$ is the dark matter phase space distribution which depends on the 3-momentum ${\bf p}_\chi$ (more precisely, it depends only on its modulus $p_\chi$ by isotropy), $E_\chi$ is the corresponding energy, and $C[f_\chi]$ is the collision operator governing the interactions between $\chi$ and the other particle species.

In the coannihilation phase at small mixing angle, as well as in the coscattering phase, the only processes contributing sizeably to $C[f_\chi]$ are inelastic scatterings off the thermal bath,
\be\label{eq:scattering}
\chi X\;\leftrightarrow\;\psi X'\,,
\ee
where $X$ and $X'$ are Standard Model particles and $\psi$ can be either $\psi^0$, $\psi^+$ or $\psi^-$. Therefore, the collision operator $C[f_\chi]$ is given by
\be\label{eq:collision}
C[f_\chi]=\sum_{\psi, X, X'}\frac{1}{2}\int \wt{\d p}_X \wt{\d p}_{X'}\wt{\d p}_\psi\;\delta^{(4)}\left(\sum p_i\right)|\ol M|^2\;\left(f_\psi f_{X'}-f_\chi f_X\right)
\ee
with $\ol M$ the matrix element for the process \eqref{eq:scattering}, suitably summed and averaged over final and initial internal degrees of freedom, and $\wt{\d p}_i$ the Lorentz-invariant phase space measure
\be
\wt{\d p}_i=\frac{\d^3 p_i}{(2\pi)^3\,2\,E_i}\,.
\ee
We can safely assume that $X$ and $X'$ are in chemical and kinetic equilibrium throughout. Moreover, we can assume that at least kinetic equilibrium is maintained for $\psi$ because of efficient elastic and inelastic scattering processes with the thermal bath. However, we allow for the integrated number density 
\be
n_\psi(t)=4\pi g_\psi\,\int\frac{p_\psi^2\d p_\psi}{(2\pi)^3}f_\psi(p_\psi,t)
\ee
to deviate from its equilibrium value, which in the Maxwell-Boltzmann approximation ($f^{\rm eq}_\psi=e^{-E_\psi/T}$, neglecting quantum statistical factors) is given by
\be
n_\psi^{\rm eq}=\frac{g_\psi\,T\,m_\psi^2}{2\pi^2}K_2\left(\frac{m_\psi}{T}\right)\,.
\ee 
Hence for $\psi$ we use the standard ansatz
\be
f_\psi(p_\psi,t)=f_\psi^{\rm eq}(p_\psi,t)\frac{n_\psi(t)}{n_\psi^{\rm eq}(t)}\,,
\ee
where $n_\psi(t)$ is an unknown function to be determined. Detailed balance allows to replace, in Eq.~\eqref{eq:collision},
\be
f_\psi f_{X'}=f_\psi^{\rm eq}f_{X'}^{\rm eq}\frac{n_\psi}{n_\psi^{\rm eq}}\,\rightarrow\,f_\chi^{\rm eq} f_{X}^{\rm eq}\frac{n_\psi}{n_\psi^{\rm eq}}
\ee
so that the collision term becomes
\be\begin{split}\label{eq:collision2}
\frac{1}{E_\chi} C[f_\chi]=&\;\sum_{\psi, X, X'}\left(f_\chi^{\rm eq}\frac{n_\psi}{n_\psi^{\rm eq}}-f_\chi\right)\frac{1}{2 E_\chi}\int \wt{\d p}_X \wt{\d p}_{X'}\wt{\d p}_\psi\;\delta^{(4)}\left(\sum p_i\right)|\ol M|^2\;f^{\rm eq}_X\\
=&\;\sum_{\psi, X, X'}\left(f_\chi^{\rm eq}\frac{n_\psi}{n_\psi^{\rm eq}}-f_\chi\right)\frac{1}{E_\chi}\int \frac{p_X\d p_X\d\phi_X\d\cos\theta_X}{(2\pi)^3}e^{-E_X/T}\;w(s)\,.\\
\end{split}
\ee
Here we have again neglected quantum statistical factors, setting  $f_{X}^{\rm eq}=e^{-E_X/T}$.
The function $w(s)$ \cite{Srednicki:1988ce} is related to the unpolarized cross-section $\sigma(s)$ by
\be
w(s)=\frac{1}{2}\sqrt{s-(m_\chi+m_X)^2}\sqrt{s-(m_\chi-m_X)^2}\sigma(s)\,,
\ee
and can be obtained from the matrix element by integrating over the final state solid angle:
\be
w(s)=\frac{\sqrt{s-(m_\psi+m_{X'})^2}\sqrt{s-(m_\psi-m_{X'})^2}}{128\pi^2\,s}\int|\ol M|^2\,\d\Omega_\psi\,.
\ee
By changing integration variables from $\cos\theta_X$ to $s=m_\chi^2+m_X^2 + 2 E_\chi E_X-2 p_\chi p_X\cos\theta_X$, and carrying out the $\phi_X$ and $p_X$ integrations, Eq.~\eqref{eq:collision2} becomes 
\be
\frac{1}{E_\chi} C[f_\chi]=\sum_{\psi, X, X'}\left(f_\chi^{\rm eq}\frac{n_\psi}{n_\psi^{\rm eq}}-f_\chi\right)\wt C(T, p_\chi)
\ee
where
\be\label{eq:collision3}
\begin{split}
\wt C=&\;\frac{T}{8\pi^2\,E_\chi p_\chi}\int\d s\;w(s)\left(e^{-E_-(s)/T}-e^{-E_+(s)/T}\right)\,,\\
E_\pm(s)=&\;\frac{s-m_\chi^2-m_X^2}{2 \,m_\chi^2}\left(E_\chi\pm p_\chi\sqrt{1-4\frac{m_\chi^2 m_X^2}{(s-m_\chi^2-m_X^2)^2}}\right)\,.
\end{split}
\ee

The dominant contribution to scattering in the singlet-triplet model is given by $W$-mediated processes where $X$ and $X'$ are relativistic Standard Model fermions.\footnote{See the Appendix for a more detailed discussion.} Neglecting Standard Model fermion masses one obtains
\be\begin{split}\label{eq:wfermions}
w(s)=&\frac{g_2^4\theta^2}{64\pi}\left(s-m_\psi^2\right)\Biggl(\frac{1}{s}+\frac{2}{m_W^2}+\frac{m_W^2-(m_\psi-m_\chi)^2}{(s-m_\psi^2)(s-m_\chi^2)+s m_W^2}\\
&+\frac{2s+2 m_W^2-(m_\psi-m_\chi)^2}{(s-m_\psi^2)(s-m_\chi^2)} \log \frac{s m_W^2}{(s-m_\psi^2)(s-m_\chi^2)+s m_W^2}\Biggr)\,,
\end{split}
\ee
where $g_2$ is the $\SU{2}$ gauge coupling and $\theta$ is the singlet-triplet mixing angle. Summing over the possible channels, i.e.~setting $\psi=\psi^+$ and $X=\ell^+$, $\nu_\ell$, $u$, $\bar d$, $c$, $\bar s$ (where $\ell$ is any lepton) or the respective  charge-conjugate states, gives an overall factor of $36$ in the final expression for the collision term. The contribution of third-generation quarks will be Boltzmann-suppressed and therefore subdominant, since the top quark is non-relativistic at the temperatures of interest, provided that $m$ is below the TeV scale. Other subdominant contributions come from processes where $X$ and $X'$ are electroweak gauge bosons and Higgs bosons, as detailed in the Appendix.

We note that, if $\chi$ were guaranteed to be in kinetic equilibrium such that $f_\chi(p_\chi,t)=f_\chi^{\rm eq}(p_\chi,t)\frac{n_\chi(t)}{n_\chi^{\rm eq}(t)}$,  Eq.~\eqref{eq:fullBoltzmann} could be further simplified by integrating over ${\bf p}_\chi$ on both sides, thus obtaining an ordinary differential equation for the integrated number density $n_\chi(t)$, as in the standard formalism in usual coannihilation scenarios. However, in the coscattering phase of our model, kinetic equilibrium is lost at the same time as chemical equilibrium (and due to the decoupling of the same process), hence we need to solve Eq.~\eqref{eq:fullBoltzmann} for all momentum modes separately, even if the final quantity we are interested in is the overall $\chi$ abundance.

To solve  Eq.~\eqref{eq:fullBoltzmann}, we integrate over the solid angle of ${\bf p}_\chi$ and change variables from $(t, p_\chi)$ to $\left(x=\frac{m_\chi}{T(t)}, q=\frac{p_\chi}{T(t)}\right)$. In terms of these variables,
\be
p_\chi\frac{\partial}{\partial p_\chi}=q\frac{\partial}{\partial q}\,,\qquad
\frac{\partial}{\partial t}=\frac{H}{1-x\,\delta_g(x)}\left(x\frac{\partial}{\partial x}+q\frac{\partial}{\partial q}\right)\,,
\ee
where we have defined
\be
\delta_g(x)=-\frac{1}{3\,g_*}\frac{\partial g_*}{\partial x}\,.
\ee

The Boltzmann equation Eq.~\eqref{eq:fullBoltzmann} becomes
\be\label{eq:fullBoltzmann2}
\left(\partial_x+\delta_g(x)\,q\partial_q\right)f_\chi(x,q)=B(x, q)\,\left(f_\chi^{\rm eq}(x,q)\frac{n_{\psi}(x)}{n_\psi^{\rm eq}(x)}-f_\chi(x,q)\right)
\ee
where
\be\label{eq:defB}
B(x,q)=\frac{1-x\,\delta_g(x)}{Hx}\wt C(x,q)\,.
\ee

On each of the characteristic curves
\be\label{eq:charact}
q_\rho(x) = \rho\,\exp\left(\int_{x_0}^x \delta_g(y)\d y\right)
\ee
Eq.~\eqref{eq:fullBoltzmann2} becomes an ordinary differential equation
\be\label{eq:BEfam}
\frac{\d}{\d x} f_\chi \left(x,\,q_\rho(x)\right)=B \left(x,\,q_\rho(x)\right)  \left(f_\chi^{\rm eq}(x,q_\rho(x))\frac{n_\psi(x)}{n_\psi^{\rm eq}(x)}-f_\chi(x,q_\rho(x))\right)\,.
\ee
Here the family of characteristics is parameterized by $\rho$ such that $q_\rho(x_0)=\rho$. The initial conditions for these ODEs are obtained by imposing that at early times, i.e.~at some $x_0\approx 1$, $\chi$ is in equilibrium:
\be
f_\chi(x_0, q)=f_\chi^{\rm eq}(x_0,q)=e^{-\sqrt{x_0^2+q^2}},\text{ hence }\left.f_\chi(x, q_\rho(x))\right|_{x=x_0}=f^{\rm eq}_\chi(x_0,\rho)\,.
\ee

The solution of Eq.~\eqref{eq:BEfam} subject to these initial condition is
\be\label{eq:solutionfchi}
\begin{split}
f_\chi(x, q_\rho(x))=&\;f_\chi^{\rm eq}(x_0, \rho)\exp\left(-\int_{x_0}^{x} B(y, q_\rho(y))\,\d y\right)\\
&+\int_{x_0}^{x} \frac{n_\psi(y)}{n_\psi^{\rm eq}(y)} f_\chi^{\rm eq}(y, q_\rho(y))\,B(y,q_\rho(y))\exp\left(-\int_y^{x} B(z, q_\rho(z))\,\d z\right)\,\d y\,.
\end{split}
\ee
The right-hand side of Eq.~\eqref{eq:solutionfchi} depends on the $\psi^+$ abundance $n_{\psi}(x)$, which is itself given by the solution of a Boltzmann equation whose collision operator depends on $f_\chi$. At sufficiently large temperatures, or equivalently at small $x$, one may set $n_{\psi}(x)\approx n^{\rm eq}_{\psi}(x)$ and use Eq.~\eqref{eq:solutionfchi} to find the solution for $f_\chi(x)$. In general, this is no longer true at larger $x$, so the coupled system of Boltzmann equations for the $\chi$ modes as well as for the momentum-integrated $\psi$ number densities needs to be solved.

The Boltzmann equations governing the $\psi$ abundances are most conveniently expressed in terms of the dimensionless quantities
\be
Y_\psi(x)=\frac{n_\psi(x)}{\hat s(x)}\,,
\ee
where
\be
\hat s=\frac{2\pi^2}{45}g_*\,T^3
\ee
is the entropy density.\footnote{We use the symbol $\hat s$ to distinguish the entropy density from the Mandelstam variable $s$ introduced earlier. Since we are only concerned with temperatures far above neutrino decoupling, we do not distinguish between effective degrees of freedom contributing to entropy and energy.} Following the steps in the standard coannihilation formalism \cite{Edsjo:1997bg} one obtains 
\be\begin{split}\label{eq:Boltzmannsystem}
\frac{\d}{\d x} Y_{\psi^+}=\frac{1}{3Hx}\frac{\d \hat s}{\d x}\Biggl[\sum_{X}\Biggl(&\langle\sigma v_{\psi^+\psi^0\into X}\rangle\left(Y_{\psi^+} Y_{\psi^0}-(Y^{\rm eq}_{\psi})^2\right) \\
&+\left(\langle\sigma v_{\psi^+\psi^-\into X}\rangle+\langle\sigma v_{\psi^+\psi^+\into X}\rangle\right)\left(Y_{\psi^+}^2-(Y^{\rm eq}_{\psi})^2\right)\Biggr)\\
 -\sum_{XX'}\Biggl(&\langle\sigma v_{\psi^0 X\into\psi^+ X'}\rangle\,Y^{\rm eq}_X\left(Y_{\psi^0}-Y_{\psi^+}\right) 
 +\frac{1}{\hat s}\int\frac{\d^3 q}{(2\pi)^3} \wt C \Bigl(f_\chi^{\rm eq}\,\frac{Y_{\psi^+}}{Y_\psi^{\rm eq}}-f_\chi\Bigr)\Biggr)\Biggr]\,,\\
 \frac{\d}{\d x} Y_{\psi^0}=\frac{1}{3Hx}\frac{\d \hat s}{\d x}\Biggr[\sum_{X}\Biggl(&2\langle\sigma v_{\psi^+\psi^0\into X}\rangle\left(Y_{\psi^+} Y_{\psi^0}-(Y^{\rm eq}_{\psi})^2\right)+\langle\sigma v_{\psi^0\psi^0\into X}\rangle\left(Y_{\psi^0}^2-(Y^{\rm eq}_{\psi})^2\right)\Biggr)\\
 +\sum_{XX'}2\,\langle&\sigma v_{\psi^0 X\into\psi^+ X'}\rangle\,Y^{\rm eq}_X\left(Y_{\psi^0}-Y_{\psi^+}\right) \Biggr]\,.\end{split}
\ee
Here we have set $Y_{\psi^+}^{\rm eq}=Y_{\psi^0}^{\rm eq}\equiv Y_\psi^{\rm eq}$, neglecting the small difference between $m_{\psi^0}$ and $m_{\psi^+}$. It is understood that $Y_{\psi^-}=Y_{\psi^+}$, and the factors of $2$ in the fourth and fifth line of Eqns.~\eqref{eq:Boltzmannsystem} are to account for terms where $\psi^-$ appears instead of $\psi^+$. The thermally averaged annihilation cross-sections are defined as usual:
\be
\langle\sigma v_{ij\into k\ell}\rangle=\frac{g_i g_j}{512\pi^6 n_i^{\rm eq} n_j^{\rm eq}} T \int\d s\,\d\Omega_k\,\frac{p_{ij} p_{k\ell}}{\sqrt{s}}|\ol M|^2 K_1\left(\frac{\sqrt{s}}{T}\right)\,,
\ee
\be
p_{ij}=\frac{\sqrt{s-(m_i-m_j)^2}\sqrt{s-(m_i+m_j)^2}}{2\sqrt{s}}\,.
\ee
Finally, $\langle\sigma v_{\psi^0 X\into\psi^+ X'}\rangle$ conversions are always efficient, which allows to work with a single effective $\psi$ abundance $Y_\psi=Y_{\psi^+}=Y_{\psi^0}$, similarly as in the standard coannihilation formalism. Summing the Boltzmann equations for  $Y_{\psi^+}$, $Y_{\psi^-}$ and $Y_{\psi^0}$, the $\psi\into\psi$ conversion terms cancel and one obtains
\be
\begin{split}\label{eq:Boltzmannpsi}
\frac{\d}{\d x} & Y_{\psi}=\frac{1}{3}\frac{1}{3Hx}\frac{\d \hat s}{\d x}\Biggl[\frac{2}{\hat s}\sum_{XX'}\int\frac{\d^3 q}{(2\pi)^3} \wt C \Bigl(f_\chi^{\rm eq}\,\frac{Y_{\psi}}{Y_\psi^{\rm eq}}-f_\chi\Bigr)\\
&+\sum_X \Biggl(4\langle\sigma v_{\psi^+\psi^0\into X}\rangle+2\langle\sigma v_{\psi^+\psi^-\into X}\rangle +2\langle\sigma v_{\psi^+\psi^+\into X}\rangle+\langle\sigma v_{\psi^0\psi^0\into X}\rangle\Biggr)\left(Y_{\psi}{}^2 -(Y^{\rm eq}_{\psi})^2\right)\Biggr]\,.
\end{split}
\ee

Eqns.~\eqref{eq:BEfam} and \eqref{eq:Boltzmannpsi} constitute a coupled system of ordinary differential equations which we need to solve, subject to the initial conditions at some suitably small $x_0$
\be
f_\chi(x_0, q_\rho(x_0))=f_\chi^{\rm eq}(x_0,\rho)\,,\qquad Y_{\psi}(x_0)=Y_\psi^{\rm eq}(x_0)\,.
\ee

For a numerical solution of this system, we discretize $q$ and work with $N=30$ characteristic curves. The initial values of $q$ at $x=x_0=1$ are chosen to lie between $\rho=0$ and $\rho=50$, at suitable points for efficient computation of the $q$-integral in Eq.~\eqref{eq:Boltzmannpsi} via $N$-point Gauss-Legendre quadrature. Since $\delta_g$ is numerically small, the range of $q$ covered by these curves diminishes by only about $10\%$ between $x=1$ and $x=400$, see Eq.~\eqref{eq:charact}. We extract the annihilation cross-sections for the $\psi$-like states from {\tt micrOMEGAs} \cite{Belanger:2001fz}. Using {\tt CalcHEP} \cite{Belyaev:2012qa} to compute the subdominant contributions to the $\chi X\into\psi X'$ scattering cross-sections (i.e.~all contributions except those involving light fermions, for which we have the analytic expression Eq.~\eqref{eq:wfermions}), we find that including these corrections changes the results at the sub-percent level for $M=500$ GeV and by $< 2\%$ for $M=1000$ GeV.
To speed up the computation, we impose that $\psi$ remains in equilibrium between $x_0=1$ and an intermediate temperature $x_{\rm int}=15$, which allows to calculate $f_\chi(x_{\rm int}, q_\rho(x_{\rm int}))$ directly from Eq.~\eqref{eq:solutionfchi}. For $x>x_{\rm int}$ the full system of 31 differential equations is then solved using a standard adaptive fourth-order Runge-Kutta method. 

\begin{figure}
 \begin{tabular}{cc}
  \includegraphics[width=.5\textwidth]{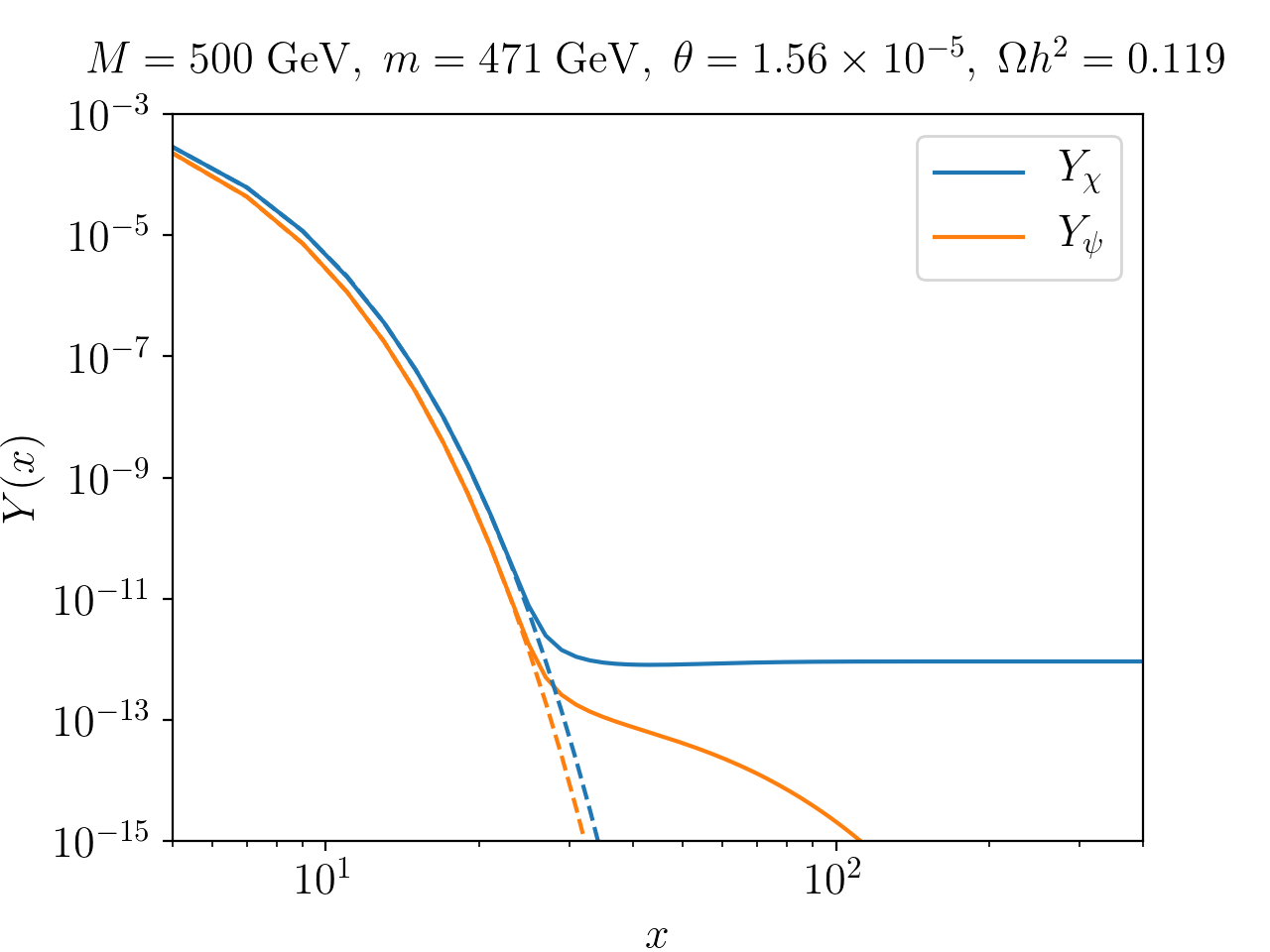} &  \includegraphics[width=.5\textwidth]{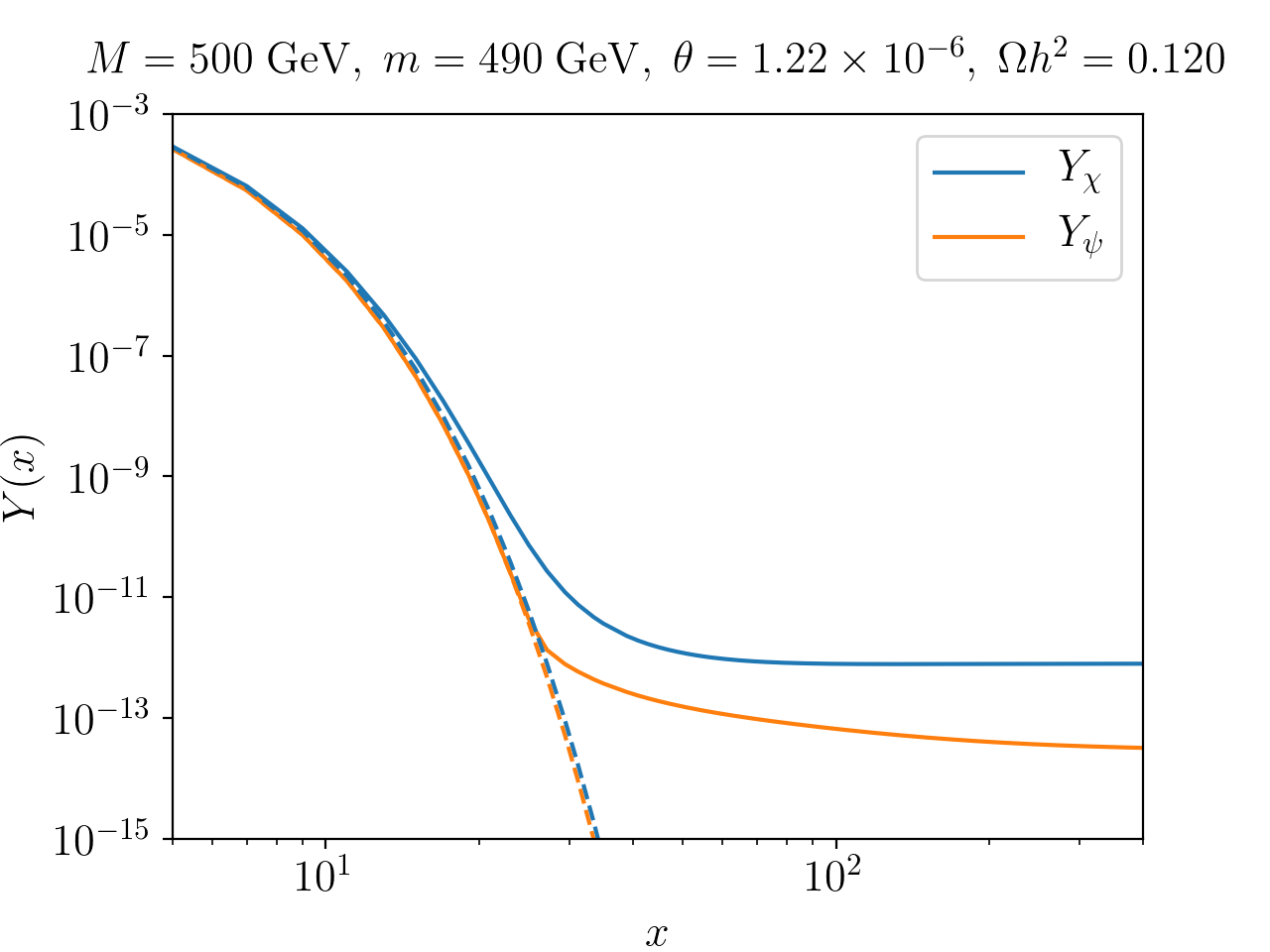}
 \end{tabular}\caption{The abundances $Y_\chi(x)$ (solid blue) and $Y_\psi(x)$ (solid orange) as a function of $x=m/T$ for $M=500$ GeV. The dashed lines correspond to the respective equilibrium abundances. Left panel: At $m=471$ GeV and $\theta=1.56\times 10^{-5}$, the model is in the coannihilation phase. Right panel: A point in the coscattering phase at $m=490$ GeV and $\theta=1.22\times 10^{-6}$.}
\label{fig:Omegah2}
\end{figure}

Fig.~\ref{fig:Omegah2} shows the evolution of $Y_\chi$ and $Y_\psi$ as a function of $x$ for two representative points, one in the coannihilation phase and the other in the coscattering phase. In the coannihilation phase, the $\chi$ number density evolves along with the $\psi$ number density, departing from equilibrium around $x=x_f\approx 25$, where $\psi\psi\into X X'$ annihilations freeze out. At larger $x$ it becomes approximately constant, while $Y_\psi$ continues to decline due to $\psi X\into \chi X'$ conversion which is still efficient (and energetically favored over the reverse process).
In the coscattering phase, the dark matter abundance departs from its equilibrium value at relatively small $x$, whereas the $\psi$ abundance remains in equilibrium until $x=x_f$. Its slow decline at larger $x$ is due to the now marginally efficient conversion processes. Eventually all remaining $\psi$-like states will decay into $\chi$, but since these decays are three-body phase-space suppressed and therefore happen on a much larger timescale, we do not take them into account here. (The final relic abundance is nevertheless calculated from the total number density $Y_\chi+Y_{\psi^0}+Y_{\psi^+}+Y_{\psi^-}=Y_\chi+3\,Y_\psi$.)

As is evident from the right panel of Fig.~\ref{fig:Omegah2}, the freeze-out of the scattering processes happens gradually, the deviation from the equilibrium number density being noticeable long before the number density ultimately settles and becomes constant. This is in marked contrast to the freeze-out of $\psi$ annihilations which happens relatively abruptly. 

\begin{figure}
 \begin{tabular}{cc}
  \includegraphics[width=.5\textwidth]{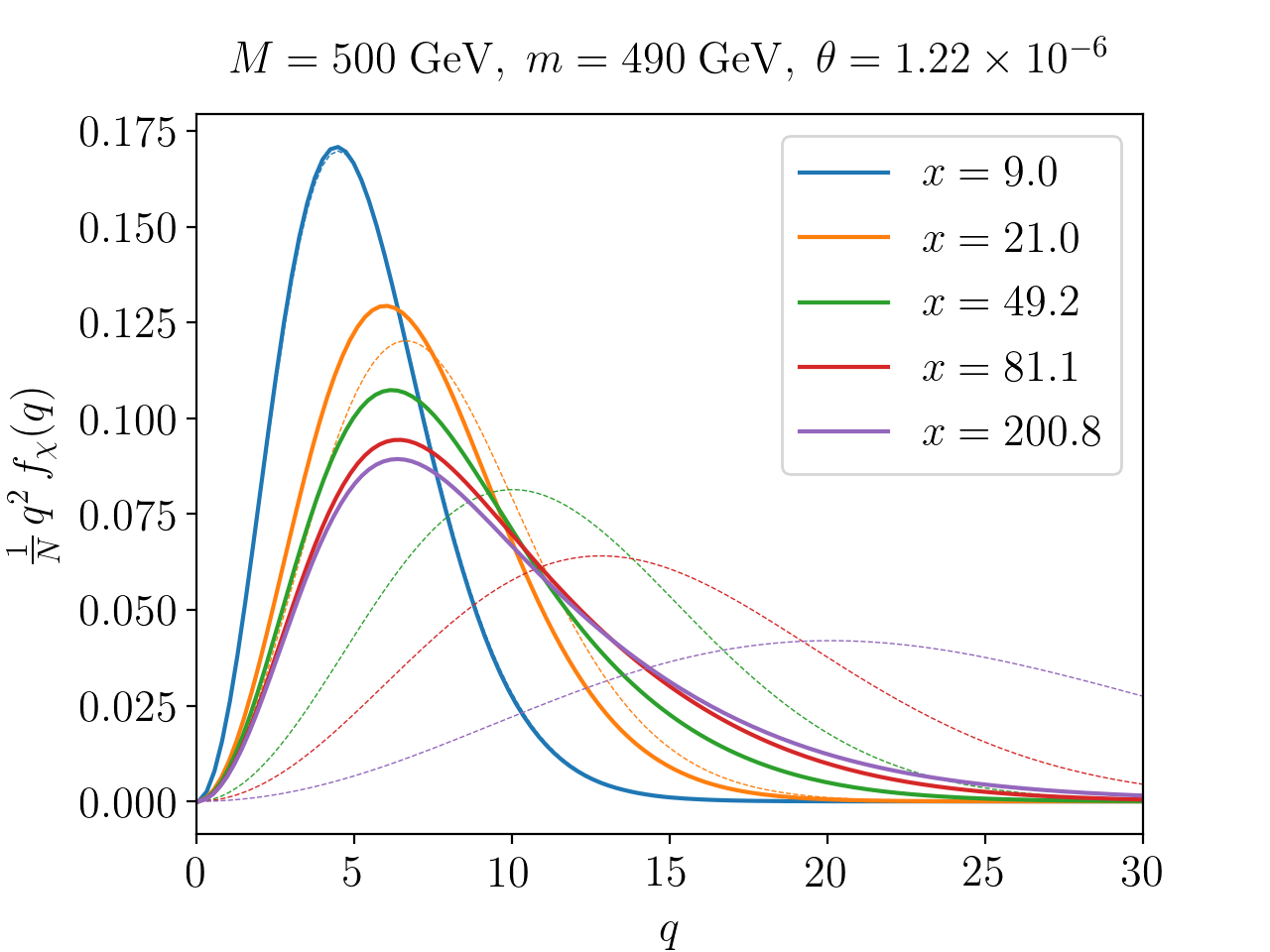} &  \includegraphics[width=.5\textwidth]{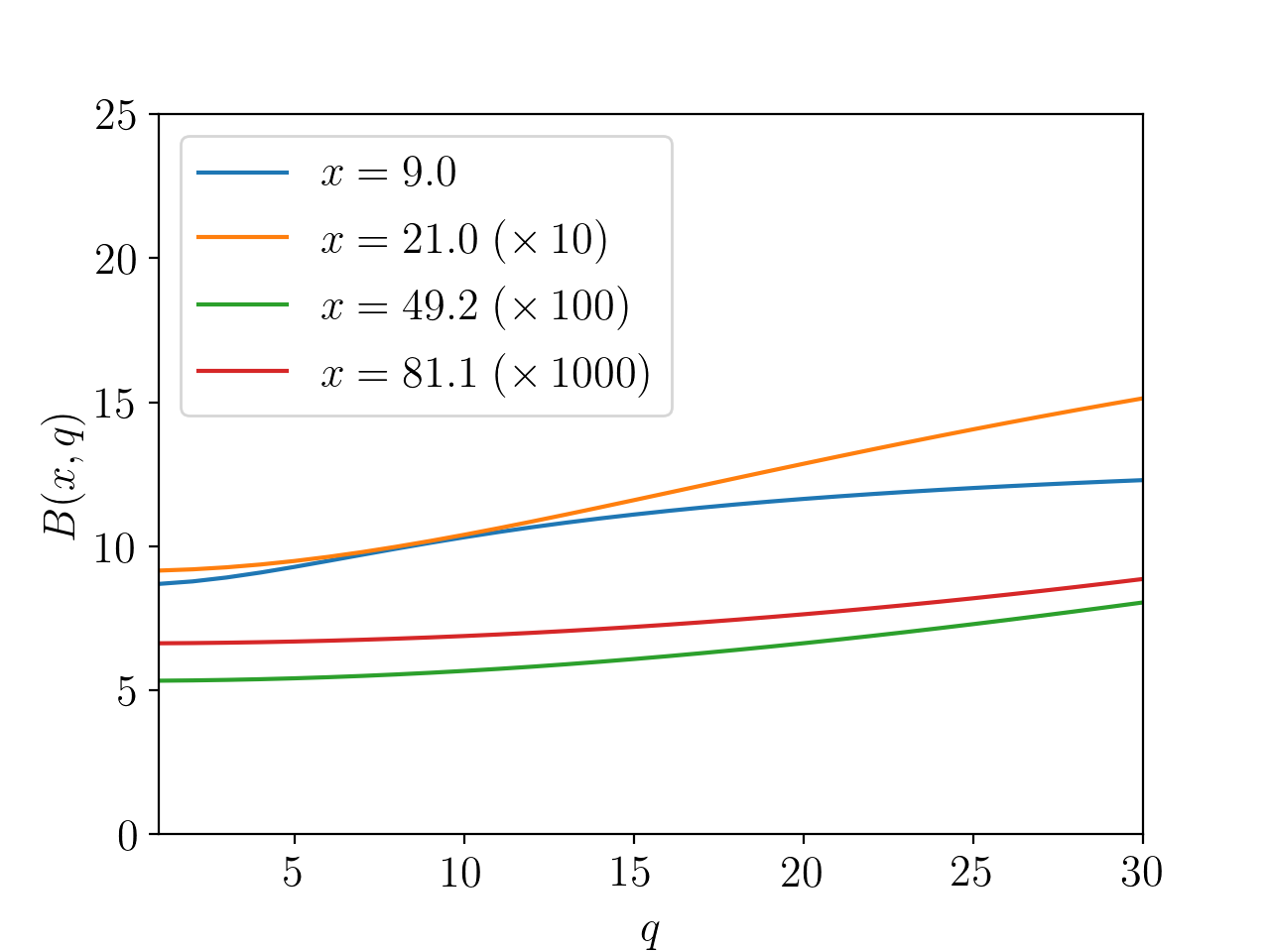}
 \end{tabular}\caption{Left panel, solid lines: the evolution of the normalized momentum mode distribution $\frac{1}{N} q^2 f_\chi(q)$ in the coscattering phase, where $N=\int\d q\,q^2f_\chi(q)$. Dashed lines: the corresponding normalized equilibrium distributions  $\frac{1}{N}q^2 f^{\rm eq}_\chi(q)$. Right panel: the rescaled collision rate $B(x,q)$ defined in Eq.~\eqref{eq:defB} for various $x$. The model parameters are the same as for the right panel of Fig.~\ref{fig:Omegah2}. }
\label{fig:modes}
\end{figure}

The left panel of Fig.~\ref{fig:modes} shows the shape of the momentum distributions $q^2 f_\chi(q)$ at several $x$ in the coscattering phase. One observes a marked departure from kinetic equilibrium starting at $x\gtrsim 10$, with the lower momentum modes decoupling earlier. This can also be seen from the right panel, which shows the rescaled collision operator $B$ of Eq.~\eqref{eq:fullBoltzmann2} for the same data point as a function of $q$; one observes that the effective collision rate increases with  $q$. When the momentum distribution is finally frozen (which is the case at $x\gtrsim 100$ for this choice of parameters), it bears little resemblance to a Maxwell-Boltzmann equilibrium distribution, leaning noticeably towards the higher modes to the right of the maximum. Consequently, the final relic abundance cannot be calculated reliably with the momentum-integrated Boltzmann equations by setting $f_\chi(q,x)\propto f_\chi^{\rm eq}(q,x)$. In the coscattering phase, using this simplification may change the predicted result for the relic density by factor of a few, and its use is justified only in the coannihilation phase. 

\begin{figure}
 \begin{tabular}{cc}
\includegraphics[width=.5\textwidth]{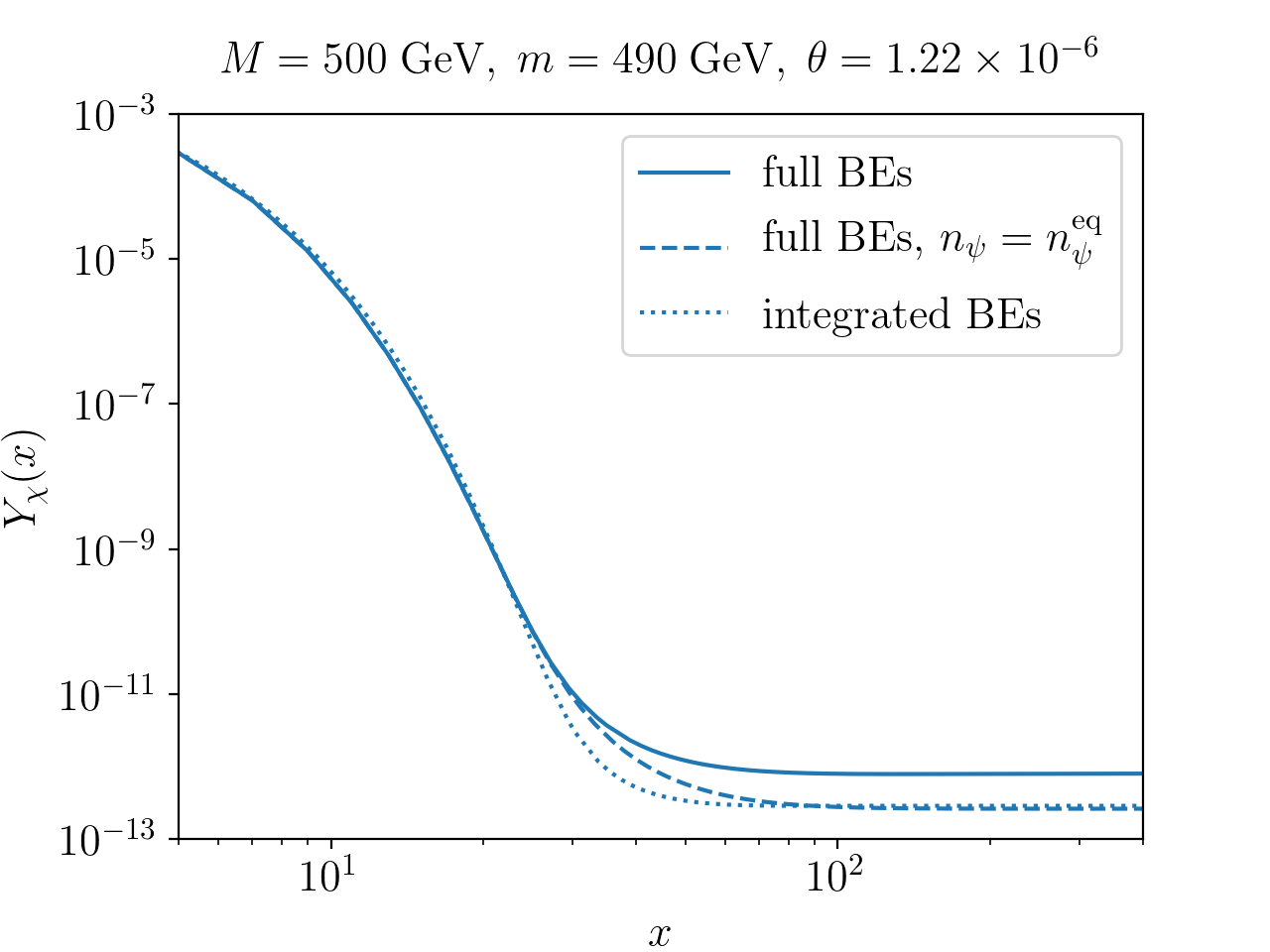} &  \includegraphics[width=.5\textwidth]{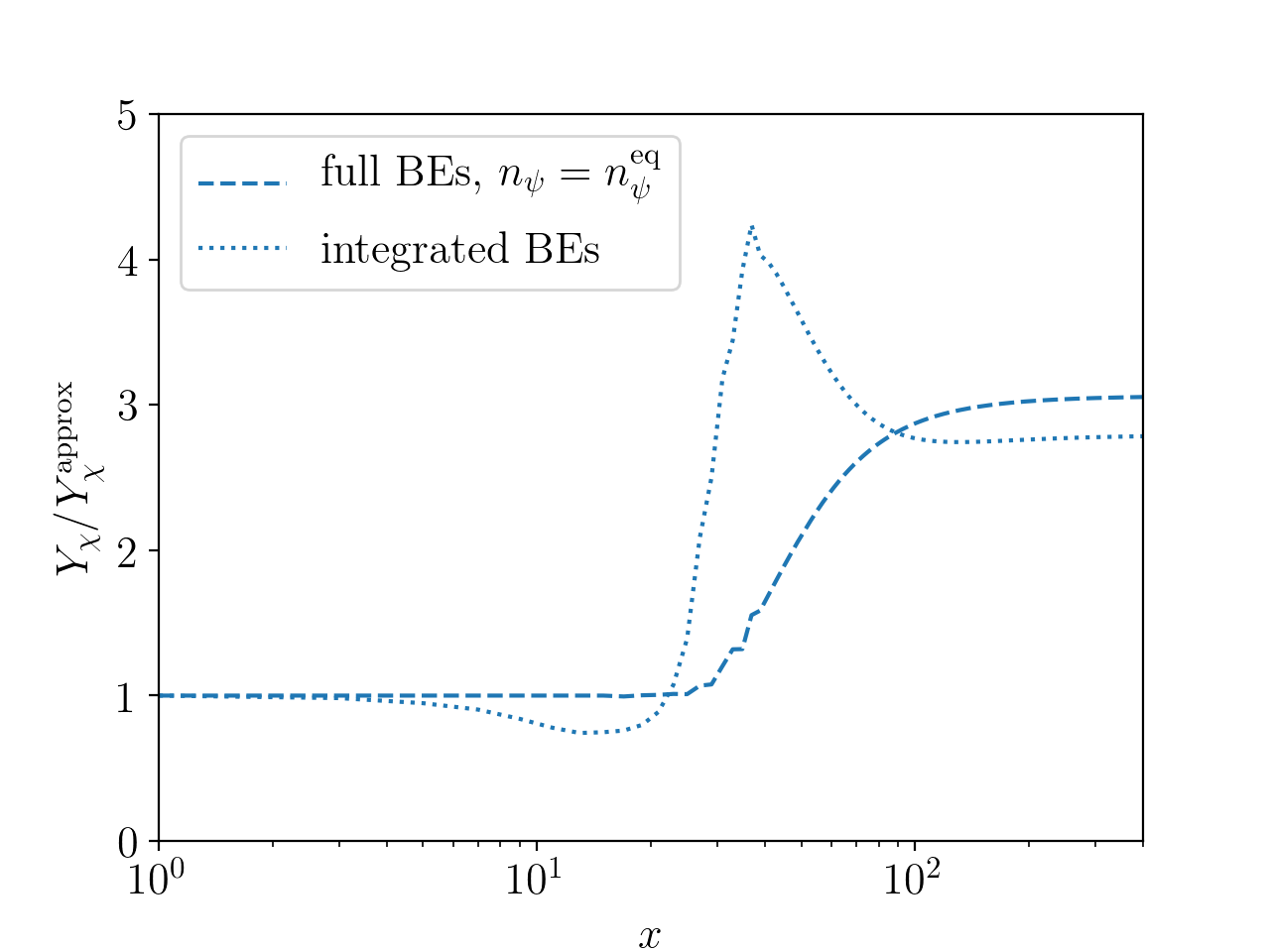}
\end{tabular}
\caption{In the left panel, the abundance $Y_\chi$ for a point in the coscattering phase, calculated using the formalism presented in the text (solid curve); assuming that $\psi$ remains in equilibrium throughout, i.e.~$n_\psi=n_\psi^{\rm eq}$ in Eq.~\eqref{eq:BEfam} (dashed curve); and calculated by imposing $f_\chi\propto f_\chi^{\rm eq}$ and solving a momentum-integrated Boltzmann equation for $Y_\chi$ similar to Eq.~\eqref{eq:Boltzmannpsi} (dotted curve). In the right panel, the ratio $Y_\chi/Y_\chi^{\rm approx}$ between the actual $\chi$ abundance and the abundance obtained with these two approximations.}
\label{fig:approx}
\end{figure}

This is illustrated in Fig.~\ref{fig:approx}, where we plot $Y_\chi$ along with the prediction for $Y_\chi$ obtained by two approximations. The first approximation (the dotted curve) is to impose $f_\chi(q,x)\propto f_\chi^{\rm eq}(q,x)$ and to solve a momentum-integrated Boltzmann equation not just for $\psi$ but also for $\chi$. The second approximation (the dashed curve) is to solve the momentum-dependent Boltzmann equation but to impose $n_\psi(x)=n_\psi^{\rm eq}(x)$ for all $x$. Evidently, the use of neither of these approximations is justified in the coscattering phase. We have verified, however, that the $\chi$ abundance can be reliably calculated using the momentum-integrated Boltzmann equations in the coannihilation phase, as expected.

Thus, in the singlet-triplet model, early kinetic decoupling leaves a significant imprint on the final relic abundance. This is in contrast to the earlier study \cite{Garny:2017rxs}, which found numerically similar results for the relic densities predicted with either the full Boltzmann equations or the integrated ones, with a difference  of the order of only 10\%. However, the model considered in that study differs from ours in one crucial aspect, namely, the importance of decays and inverse decays $\psi\,\leftrightarrow\, X\chi$. In our model, these processes are phase-space suppressed because the mass difference $M-m$ is always below $m_W$, whereas in the model of \cite{Garny:2017rxs} the mediator particle (the analogue of our $\psi$) can decay into $\chi$ and an on-shell $b$ quark, rendering decays and inverse decays efficient until rather late times and thus helping to re-establish approximate kinetic equilibrium.

\begin{figure}
\begin{tabular}{cc}
  \includegraphics[width=.5\textwidth]{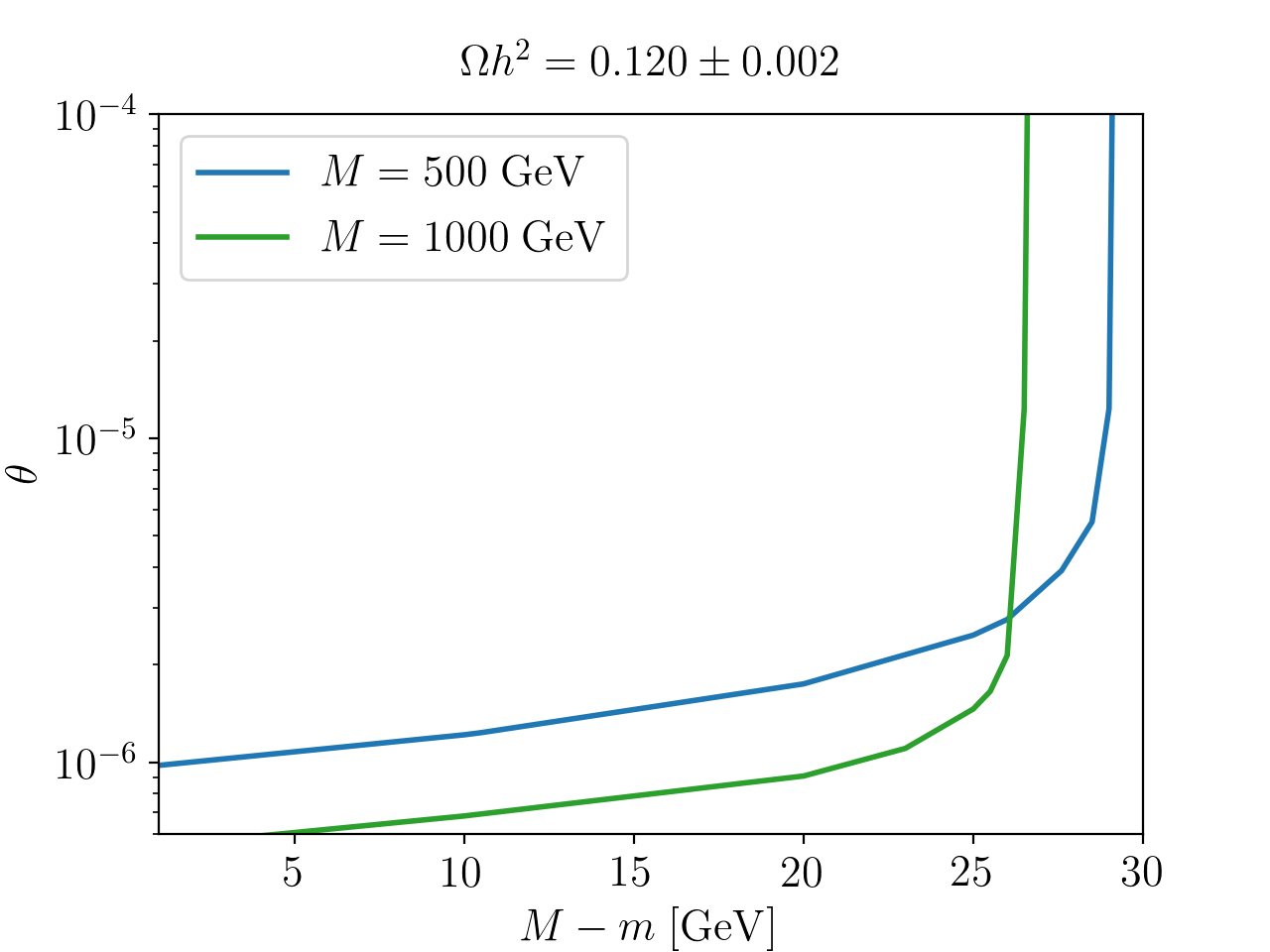} &  \includegraphics[width=.5\textwidth]{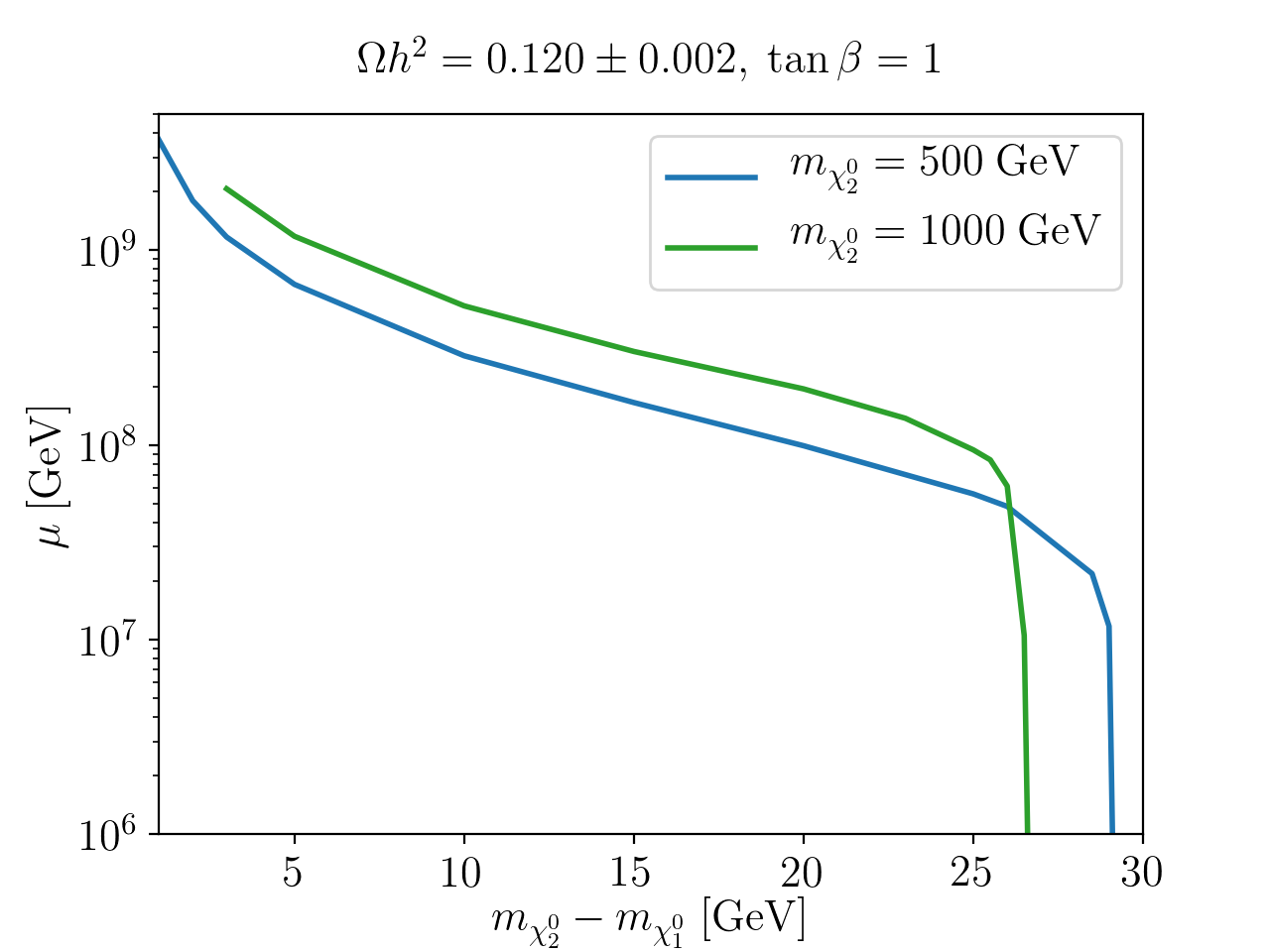} \end{tabular}
\caption{Left panel: The observed dark matter relic density is reproduced on the blue curve for $M=500$ GeV and on the green curve for $M=1000$ GeV, shown in the plane of the mass difference $M-m$ (for positive $m$) and the mixing angle $\theta$. The vertical sections on the right of the curves correspond to the coannihilation phase (which extends to larger mixing angles), the diagonal sections on the bottom left represent the coscattering phase. In the regions on the left of the curves, the $\chi$ abundance is too low to account for all of the dark matter in the universe, whereas in the regions on the right the model predicts too much dark matter. Right panel: In split supersymmetry,  the same constraint expressed in the plane of $m_{\chi^0_2}-m_{\chi^0_1}$ and the $\mu$ parameter at $\tan\beta=1$, assuming $\mu>0$.}
\label{fig:transition}
\end{figure}

We illustrate the transition between the coannihilation and the coscattering phase in the singlet-triplet model in Fig.~\ref{fig:transition}, where the curve indicates the parameters for which the observed dark matter relic density $\Omega h^2=0.120$ \cite{Aghanim:2018eyx} is reproduced. Here we are assuming $m>0$, hence we can identify $m_\chi= m$ and $m_{\psi^\pm}= m_{\psi^0}= M$. In the coannihilation phase, the predicted relic density is almost independent of the mixing angle but depends sensitively on the mass difference $M-m$, whereas in the coscattering phase, it is the mixing angle which determines at what time the dark matter density departs from its equilibrium value.

\section{Split supersymmetry with a bino LSP}
\label{sec:split}

The results of the previous section are valid for a general singlet-triplet model. However, when supposing that the UV completion is split supersymmetry, we can use them to obtain information on the value of the $\mu$ parameter. This is because the split supersymmetry couplings $\tilde g_{u,d}$ and $\tilde g_{u,d}'$ are given by the electroweak gauge couplings and by $\tan\beta$ at the scale $M_{\rm SUSY}$,
\be\label{eq:tildedmatching}
\tilde g_u=g_2\sin\beta \,,\quad \tilde g_d= g_2\cos\beta\,,\quad \tilde g_u'=\sqrt{\frac{3}{5}}\, g_1\sin\beta\,,\quad \tilde g_d'=\sqrt{\frac{3}{5}}\, g_1\cos\beta\quad\text{ at } M_{\rm SUSY}\,.
\ee
Moreover, $\tan\beta$ and $M_{\rm SUSY}$ are related by the condition
\be\label{eq:quarticmatching}
\lambda_H=\frac{1}{4}\left(\frac{3}{5}\,g_1^2+g_2^2\right)\cos^2 2\beta\quad\text{ at } M_{\rm SUSY}\,,
\ee
where $\lambda_H$ is the Standard Model Higgs quartic coupling. The couplings $\tilde g_{u,d}$ and $\tilde g_{u,d}'$  determine the Wilson coefficient $\lambda$ of Eq.~\eqref{eq:Lnonren} at the scale $|\mu|$ according the matching condition Eq.~\eqref{eq:ssmatching}. Between $|\mu|$ and the electroweak scale, $\lambda$ evolves according to its renormalization group equation (RGE), which at the one-loop level reads
\be\begin{split}
\frac{\d\lambda}{\d t} =&\; \frac{\lambda}{16\pi^2}\left(2\tr Y_e^\dag Y_e+6\,\tr Y_u^\dag Y_u+6\,\tr Y_d^\dag Y_d+\lambda_H-\frac{3}{2}g_1^2-\frac{9}{2}g_2^2\right)\,.\\
\end{split}
\ee
Finally, at the electroweak scale we have
\be\label{eq:ssmixing}
\theta=\frac{|\lambda|\, v^2}{2|\mu| (M-m)}\,.
\ee
Therefore, given either $M_{\rm SUSY}$ or $\tan\beta$, the mixing angle $\theta$ is fully determined by the bino, wino and higgsino masses $\pm m=m_{\chi^0_1}$, $M=m_{\chi^0_2}$ and $|\mu|=m_{\chi^0_{3,4}}$. 

Fixing $\tan\beta$, we can now pick any point in the space of $\theta$, $M$ and $m$ which gives rise to the observed relic density. We can then determine the corresponding $\mu$ by iteratively solving the RGEs for $\lambda$ and the Standard Model couplings
between the electroweak scale and $|\mu|$, and the split supersymmetry RGEs \cite{Giudice:2004tc} between $|\mu|$ and $M_{\rm SUSY}$, using the matching conditions Eqns.~\eqref{eq:ssmatching}, \eqref{eq:tildedmatching} and \eqref{eq:quarticmatching} and the electroweak-scale relation Eq.~\eqref{eq:ssmixing}.

The result for $\tan\beta=1$ is shown in the right panel of Fig.~\ref{fig:transition}. To obtain this curve we have used two-loop RGEs for the gauge couplings and one-loop RGEs with some dominant two-loop corrections for the others, with tree-level matching. While the precision of this calculation could be improved with state-of-the-art tools, we find that it is sufficient to estimate $\mu$ in this scenario, given that the RG evolution of the relevant couplings tends to be rather slow (a naive ``tree-level'' estimation of $\mu$, using electroweak-scale values for all couplings, is within 5\% of our loop-corrected result). For this choice of parameters, the scale $M_{\rm SUSY}$ is given by the scale where the Standard Model Higgs quartic coupling $\lambda_H$ crosses zero. One finds that $M_{\rm SUSY}$ tends to be about two orders of magnitude above $\mu$, but since the RG evolution of $\lambda_H$ is subject to large uncertainties due to the uncertainty on the top Yukawa coupling, this result should be taken with a grain of salt.

It is interesting to note that a bino-like LSP coannihilating or coscattering with a wino-like neutralino and chargino remains an open and testable option for neutralino dark matter in split supersymmetry, besides the much-studied almost-pure bino and almost-pure higgsino cases which are not testable at present colliders (we remark that a sub-TeV bino-like LSP coannihilating with a higgsino is ruled out by now, see e.g.~\cite{Krall:2017xij}). Indeed, our analysis shows that sub-TeV neutralino dark matter remains perfectly viable and necessarily comes with a coannihilation partner within the kinematic reach of the LHC, see Section \ref{sec:collider}. 

A few comments about split supersymmetry with a very large $\mu$ term are finally in order. First, it is debatable to what extent the hierarchy $|\mu|\gg |M_{1,2}|$ is natural and whether or not that constitutes a problem. One might argue that the $\mu$ parameter is allowed by supersymmetry, so whatever dynamics generates the higgsino masses need not involve supersymmetry breaking, and the mass scales $|\mu|$ and $|M_i|$ need not be correlated. However, it is well known that in split supersymmetry the gaugino masses and $\mu$ are no longer separately protected against additive renormalization. In fact, a large $\mu$ parameter induces large one-loop threshold corrections to the gaugino masses at the higgsino scale:
\be
\Delta M_1=\frac{\tilde g_u'\,\tilde g_d'}{8\pi^2}\,\mu\,\log\frac{\mu^2}{m_A^2}\,,\qquad\Delta M_2=\frac{\tilde g_u\,\tilde g_d}{8\pi^2}\,\mu\,\log\frac{\mu^2}{m_A^2}\,.
\ee
Here $m_A\approx M_{\rm SUSY}$ is the mass scale of the heavy Higgs bosons. Hence, a hierarchy $|\mu| / |M_{1,2}|\gtrsim 100$ at moderate $\tan\beta$ requires some fine-tuning of the model parameters, since large radiative corrections to the gaugino masses must be cancelled against large tree-level values. We will not attempt to quantify the required fine-tuning here given that split supersymmetry is unnatural by construction anyway (not to mention that the dark matter relic density is also rather sensitive to the precise value of the mass difference $M-m$). 

Second, note that the usual upper bound on the UV completion scale of split supersymmetry from vacuum stability \cite{Bagnaschi:2014rsa} does not apply to the case where higgsinos are heavy. This is because, in the absence of dynamical higgsinos, the renormalization group running of the Higgs quartic coupling $\lambda_H$ is not modified at the one-loop level with respect to the Standard Model. Therefore this coupling can remain positive up to high scales, as is confirmed by our numerical analysis. In particular, a mass hierarchy $M_{\rm SUSY}\gg|\mu|$ with $|\mu|\approx 10^8$ GeV is not excluded by vacuum stability.

Lastly, the gauge couplings in split supersymmetry with heavy higgsinos will not unify unless there are additional heavy states in incomplete GUT representations in the spectrum.

\section{Constraints and signatures}
\label{sec:collider}
As detailed e.g.~in \cite{Bharucha:2017ltz}, next-to-minimal dark matter at small mixing angles evades the constraints from both direct and indirect dark matter detection experiments, since the dark matter particle is essentially singlet-like.

Big bang nucleosynthesis will place constraints on the singlet-triplet model if $\chi$ and $\psi$ are almost mass degenerate, since for sufficiently small mass splittings and mixing angles, the $\psi^0\into\chi$ decay width will be highly suppressed and the $\psi^0$ lifetime can attain the range of seconds to minutes or more. Depending on the $\psi^0$ abundance after all scattering processes have frozen out, these late decays can conflict with the successful prediction for nuclear abundances. For example, taking $\theta=1.22\times 10^{-6}$, $M=500$ GeV and $m=490$ GeV, the $\psi^0$ lifetime is about $10$ s, which is still allowed by BBN constraints \cite{Kawasaki:2017bqm} given the predicted number density (see Fig.~\ref{fig:modes}). However, for lifetimes $\gtrsim 10^2$ s, the constraints on the abundance become much more severe and the model will likely be excluded. A detailed analysis of the BBN constraints is beyond the scope of this work and left for a future study.

The possible collider signatures for singlet-triplet models have been analyzed in detail e.g.~in \cite{Baer:2005jq, Han:2014xoa, Nagata:2015pra, Rolbiecki:2015gsa, Bramante:2015una, Bharucha:2018pfu, Duan:2018rls, Filimonova:2018qdc}. Therefore, here we will only briefly review the state of the most important LHC search for the model in the coscattering phase, which is the search for disappearing charged tracks. For this signature the dark matter particle $\chi$ plays practically no role, but it is the coannihilation partner $\psi$ which is being probed. In detail, the charged states $\psi^\pm$ are produced by the Drell-Yan process. For sufficiently small mixing angles $\theta \lesssim 10^{-4}$ they will preferentially decay into $\psi^0$ rather than into $\chi$; $\psi^0$ is then stable on collider scales. The mass splitting $m_{\psi^+}-m_{\psi^0}$ is induced by electroweak loops and is of the order of $160$ MeV, which implies that $\psi^\pm$ has a macroscopic lifetime because its decays are phase-space suppressed. It also implies that the only possible two-body decay is $\psi^\pm\into\psi^0 \pi^\pm$ with a very soft pion. The signature is therefore a charged ionization track left by $\psi^\pm$, with a typical length in the cm -- m range,  which at some point seemingly disappears (since the neutral $\psi^0$ is invisible and the pion is too soft to be identified).

In this channel, present data excludes triplet mass parameters below about $500$ GeV \cite{Aaboud:2017mpt, Sirunyan:2018ldc}, hence the benchmark points we have been using in Section \ref{sec:coco} are still viable, if by a small margin. At 3000 fb$^{-1}$ the projected LHC exclusion ranges up to $M\approx 900$ GeV \cite{ATLASprojected}.

\section{Summary}

We have presented a precise calculation of the dark matter relic abundance from thermal freeze-out in a scenario where the usual coannihilation formalism cannot be applied, as the relic density is set by coscattering. Our example model is singlet-triplet next-to-minimal dark matter (which can be realized in split supersymmetry), but the formalism could be applied to other similar models with little modification. In particular, it might be interesting to use it to study coscattering in models of light dark matter interacting via a light mediator. The particle abundances in the dark matter sector are governed by the full momentum-dependent Boltzmann equations, which we have cast into a form amenable to a numerical solution, properly taking into account the effects of early kinetic decoupling. 

In the singlet-triplet model, the observed dark matter density can be reproduced by coannihilations for mixing angles $\theta\gtrsim 10^{-5}$ and by coscattering for smaller mixing angles. The mass difference between the triplet-like and singlet-like states becomes smaller as the mixing angle is reduced. Our analysis also allows to chart the parameter space for bino-like dark matter in split supersymmetry with a very large $\mu$ parameter, where we find that coscattering starts setting in at values of $|\mu|\gtrsim 10^7$ GeV. While we have not undertaken a full phenomenological study of the coscattering phase, we anticipate that the most promising experimental channel to constrain it will be LHC searches for disappearing charged tracks.

\subsection*{Acknowledgments}
The author thanks A.~Bharucha and N.~Desai for numerous useful discussions about next-to-minimal dark matter models, G.~Moultaka for a Bessel function identity, and K.~Gauss for providing an incentive to start this project. This work has supported by the OCEVU Labex (ANR-11-LABX-0060) and the A*MIDEX project (ANR-11-IDEX-0001-02) funded by the "Investissements d'Avenir" French government program managed by the ANR.

\appendix

\section*{Appendix: Contributions to the collision operator for $\chi$}

In the singlet-triplet model, the dark matter particle is coupled to the Standard Model via the dimension-5 operators in Eq.~\eqref{eq:Lnonren}. After electroweak symmetry breaking, they  induce effective $hh\chi^0\chi^0$, $h h\chi^0\psi^0$, $h\chi^0\chi^0$  and $h\psi^0\chi^0$ couplings (where $h$ is the Higgs boson), and the last term gives rise to $\chi-\psi$ mixing and thus a $\chi^0\psi^\pm W_\mu$ vertex. These are the only couplings between $\chi^0$ and the Standard Model at the order $1/\Lambda$ (note that the $\chi^0\psi^0 Z_\mu$ vertex is induced by a dimension-6 operator), and therefore the only ones relevant for the collision operator $C[f_\chi]$ if $\Lambda$ is large.

Among the states in the spectrum, we have a mild mass hierarchy between the particles of the dark matter sector ($\chi^0,\psi^0,\psi^\pm$) and the Standard Model particles with masses around the electroweak scale $(t, h, Z, W)$. All other Standard Model particles can be treated as effectively massless. At the temperatures relevant to freeze-out, the dark matter sector states are of course non-relativistic, with their equilibrium number densities suppressed by Boltzmann factors $\sim e^{-m/T}$. We can therefore neglect all processes of the type $\chi^0\chi^0\into X$ and $\psi^\pm\chi^0\into X$ because they are both doubly Boltzmann-suppressed and $\Lambda$-suppressed. We can also neglect decays and inverse decays in the dark matter sector, since the mass splitting $M-m$ is below the $W$ mass, and therefore all two-body decays are kinematically forbidden. 

What remains are $2\into 2$ processes where dark matter sector particles scatter off the thermal bath. The dominant one among these is $W$-mediated $\chi^0 f\into\psi^\pm f'$ inelastic scattering where $f$ and $f'$ are light Standard Model fermions (excluding third-generation quarks), since these are highly relativistic at the relevant temperatures. Note that Higgs-mediated $\chi^0 f\into\psi^0 f$ or  $\chi^0 f\into\chi^0 f$ scattering is $m_f$-suppressed.

All processes involving  $t, h, Z$ or $W$ in the initial or final state give subdominant contributions which, however, become relatively more important as the dark matter mass increases. For dark matter masses around $500$ GeV, all these particles are already non-relativistic at $x\approx 20$, whereas for dark matter masses around $1000$ GeV, they are only on the brink of becoming non-relativistic at the temperatures of interest. For $M=500$ GeV we have checked that including the subdominant processes changes the final relic abundance by less than $1\%$, and we consequently neglect them. For $M=1000$ GeV we include the numerically most important ones among these corrections, specifically $\chi^0 t\into \psi^+ b$, $\chi^0 \bar b\into\psi^+\bar t$, $\chi^0 t\into\psi^0 t$, $\chi^0 W^+\into\psi^+Z$, $\chi^0 Z\into\psi^+W^-$, $\chi^0 W^+\into\psi^+\gamma$, $\chi^0\gamma\into\psi^+ W-$, $\chi^0 W^+\into\psi^+h$, $\chi^0 h\into\psi^+ W^-$ and $\chi^0 h\into\psi^0 h$ as well as the respective charge-conjugate processes. We find that their combined impact on the result is still at the $< 2\%$ level.


\begin{thebibliography}{99}

\bibitem{Griest:1990kh}
  K.~Griest and D.~Seckel,
  ``Three exceptions in the calculation of relic abundances,''
  Phys.\ Rev.\ D {\bf 43} (1991) 3191.

\bibitem{DAgnolo:2017dbv}
  R.~T.~D'Agnolo, D.~Pappadopulo and J.~T.~Ruderman,
  ``Fourth Exception in the Calculation of Relic Abundances,''
  Phys.\ Rev.\ Lett.\  {\bf 119} (2017) no.6,  061102
  [arXiv:1705.08450 [hep-ph]].


\bibitem{Garny:2017rxs}
  M.~Garny, J.~Heisig, B.~L\"ulf and S.~Vogl,
  ``Coannihilation without chemical equilibrium,''
  Phys.\ Rev.\ D {\bf 96} (2017) no.10,  103521
  [arXiv:1705.09292 [hep-ph]].
  
\bibitem{Gondolo:1990dk}
  P.~Gondolo and G.~Gelmini,
  ``Cosmic abundances of stable particles: Improved analysis,''
  Nucl.\ Phys.\ B {\bf 360} (1991) 145.

\bibitem{Edsjo:1997bg}
  J.~Edsj\"o and P.~Gondolo,
  ``Neutralino relic density including coannihilations,''
  Phys.\ Rev.\ D {\bf 56} (1997) 1879
  [hep-ph/9704361].
  
\bibitem{Garny:2018icg}
  M.~Garny, J.~Heisig, M.~Hufnagel and B.~L\"ulf,
  ``Top-philic dark matter within and beyond the WIMP paradigm,''
  Phys.\ Rev.\ D {\bf 97} (2018) no.7,  075002
  [arXiv:1802.00814 [hep-ph]].
  
\bibitem{Cheng:2018vaj}
  H.~C.~Cheng, L.~Li and R.~Zheng,
  ``Coscattering/Coannihilation Dark Matter in a Fraternal Twin Higgs Model,''
  JHEP {\bf 1809} (2018) 098
  [arXiv:1805.12139 [hep-ph]].
  
\bibitem{Junius:2019dci}
  S.~Junius, L.~Lopez-Honorez and A.~Mariotti,
  ``A feeble window on leptophilic dark matter,''
  JHEP {\bf 1907} (2019) 136
  [arXiv:1904.07513 [hep-ph]].
  
\bibitem{Kim:2019udq}
  H.~Kim and E.~Kuflik,
  ``Super heavy thermal dark matter,''
  arXiv:1906.00981 [hep-ph].
  
\bibitem{Bharucha:2017ltz}
  A.~Bharucha, F.~Br\"ummer and R.~Ruffault,
  ``Well-tempered n-plet dark matter,''
  JHEP {\bf 1709} (2017) 160
  [arXiv:1703.00370 [hep-ph]].
  
\bibitem{Bharucha:2018pfu}
  A.~Bharucha, F.~Br\"ummer and N.~Desai,
 ``Next-to-minimal dark matter at the LHC,''
  JHEP {\bf 1811} (2018) 195
  [arXiv:1804.02357 [hep-ph]].
 
  
\bibitem{ArkaniHamed:2004fb}
  N.~Arkani-Hamed and S.~Dimopoulos,
  ``Supersymmetric unification without low energy supersymmetry and signatures for fine-tuning at the LHC,''
  JHEP {\bf 0506} (2005) 073
  [hep-th/0405159].

\bibitem{Giudice:2004tc}
  G.~F.~Giudice and A.~Romanino,
  ``Split supersymmetry,''
  Nucl.\ Phys.\ B {\bf 699} (2004) 65
   Erratum: [Nucl.\ Phys.\ B {\bf 706} (2005) 487]
  [hep-ph/0406088].
  
\bibitem{Duch:2017nbe}
  M.~Duch and B.~Grzadkowski,
  ``Resonance enhancement of dark matter interactions: the case for early kinetic decoupling and velocity dependent resonance width,''
  JHEP {\bf 1709} (2017) 159
  [arXiv:1705.10777 [hep-ph]].
  
\bibitem{Binder:2017rgn}
  T.~Binder, T.~Bringmann, M.~Gustafsson and A.~Hryczuk,
  ``Early kinetic decoupling of dark matter: when the standard way of calculating the thermal relic density fails,''
  Phys.\ Rev.\ D {\bf 96} (2017) no.11,  115010
  [arXiv:1706.07433 [astro-ph.CO]].

\bibitem{Ibe:2012sx}
  M.~Ibe, S.~Matsumoto and R.~Sato,
  ``Mass Splitting between Charged and Neutral Winos at Two-Loop Level,''
  Phys.\ Lett.\ B {\bf 721} (2013) 252
  [arXiv:1212.5989 [hep-ph]].

\bibitem{Baer:2005jq}
  H.~Baer, T.~Krupovnickas, A.~Mustafayev, E.~K.~Park, S.~Profumo and X.~Tata,
  ``Exploring the BWCA (bino-wino co-annihilation) scenario for neutralino dark matter,''
  JHEP {\bf 0512} (2005) 011
  [hep-ph/0511034].
  
\bibitem{ArkaniHamed:2006mb}
  N.~Arkani-Hamed, A.~Delgado and G.~F.~Giudice,
  ``The Well-tempered neutralino,''
  Nucl.\ Phys.\ B {\bf 741} (2006) 108
  [hep-ph/0601041].
  
\bibitem{Ibe:2013pua}
  M.~Ibe, A.~Kamada and S.~Matsumoto,
  ``Mixed (cold+warm) dark matter in the bino-wino coannihilation scenario,''
  Phys.\ Rev.\ D {\bf 89} (2014) no.12,  123506
  [arXiv:1311.2162 [hep-ph]].
  
\bibitem{Harigaya:2014dwa}
  K.~Harigaya, K.~Kaneta and S.~Matsumoto,
  ``Gaugino coannihilations,''
  Phys.\ Rev.\ D {\bf 89} (2014) no.11,  115021
  [arXiv:1403.0715 [hep-ph]].
  
\bibitem{Bramante:2015una}
  J.~Bramante, N.~Desai, P.~Fox, A.~Martin, B.~Ostdiek and T.~Plehn,
  ``Towards the Final Word on Neutralino Dark Matter,''
  Phys.\ Rev.\ D {\bf 93} (2016) no.6,  063525
  [arXiv:1510.03460 [hep-ph]].
  
\bibitem{Yanagida:2019evh}
  T.~T.~Yanagida, W.~Yin and N.~Yokozaki,
  ``Bino-wino coannihilation as a prediction in the $E_7$ unification of families,''
  arXiv:1907.07168 [hep-ph].
  
\bibitem{Srednicki:1988ce}
  M.~Srednicki, R.~Watkins and K.~A.~Olive,
  ``Calculations of Relic Densities in the Early Universe,''
  Nucl.\ Phys.\ B {\bf 310} (1988) 693.
  
\bibitem{Belanger:2001fz}
  G.~Belanger, F.~Boudjema, A.~Pukhov and A.~Semenov,
  ``MicrOMEGAs: A Program for calculating the relic density in the MSSM,''
  Comput.\ Phys.\ Commun.\  {\bf 149} (2002) 103   [hep-ph/0112278]; ``micrOMEGAs: Version 1.3,''
  Comput.\ Phys.\ Commun.\  {\bf 174} (2006) 577 [hep-ph/0405253].
  
  
\bibitem{Belyaev:2012qa}
  A.~Belyaev, N.~D.~Christensen and A.~Pukhov,
  ``CalcHEP 3.4 for collider physics within and beyond the Standard Model,''
  Comput.\ Phys.\ Commun.\  {\bf 184} (2013) 1729
  [arXiv:1207.6082 [hep-ph]].
  
\bibitem{Aghanim:2018eyx}
  N.~Aghanim {\it et al.} [Planck Collaboration],
  ``Planck 2018 results. VI. Cosmological parameters,''
  arXiv:1807.06209 [astro-ph.CO].
  
\bibitem{Krall:2017xij}
  R.~Krall and M.~Reece,
  ``Last Electroweak WIMP Standing: Pseudo-Dirac Higgsino Status and Compact Stars as Future Probes,''
  Chin.\ Phys.\ C {\bf 42} (2018) no.4,  043105
  [arXiv:1705.04843 [hep-ph]].

\bibitem{Bagnaschi:2014rsa}
  E.~Bagnaschi, G.~F.~Giudice, P.~Slavich and A.~Strumia,
  ``Higgs Mass and Unnatural Supersymmetry,''
  JHEP {\bf 1409} (2014) 092
  [arXiv:1407.4081 [hep-ph]].
  
\bibitem{Kawasaki:2017bqm}
  M.~Kawasaki, K.~Kohri, T.~Moroi and Y.~Takaesu,
  ``Revisiting Big-Bang Nucleosynthesis Constraints on Long-Lived Decaying Particles,''
  Phys.\ Rev.\ D {\bf 97} (2018) no.2,  023502
  [arXiv:1709.01211 [hep-ph]].
  
\bibitem{Han:2014xoa}
  C.~Han, L.~Wu, J.~M.~Yang, M.~Zhang and Y.~Zhang,
  ``New approach for detecting a compressed bino/wino at the LHC,''
  Phys.\ Rev.\ D {\bf 91} (2015) 055030
  [arXiv:1409.4533 [hep-ph]].

\bibitem{Nagata:2015pra}
  N.~Nagata, H.~Otono and S.~Shirai,
  ``Probing Bino-Wino Coannihilation at the LHC,''
  JHEP {\bf 1510} (2015) 086
  [arXiv:1506.08206 [hep-ph]].

\bibitem{Rolbiecki:2015gsa}
  K.~Rolbiecki and K.~Sakurai,
  ``Long-lived bino and wino in supersymmetry with heavy scalars and higgsinos,''
  JHEP {\bf 1511} (2015) 091
  [arXiv:1506.08799 [hep-ph]].
  
\bibitem{Duan:2018rls}
  G.~H.~Duan, K.~I.~Hikasa, J.~Ren, L.~Wu and J.~M.~Yang,
  ``Probing bino-wino coannihilation dark matter below the neutrino floor at the LHC,''
  Phys.\ Rev.\ D {\bf 98} (2018) no.1,  015010
  [arXiv:1804.05238 [hep-ph]].
  
\bibitem{Filimonova:2018qdc}
  A.~Filimonova and S.~Westhoff,
  ``Long live the Higgs portal!,''
  JHEP {\bf 1902} (2019) 140
  [arXiv:1812.04628 [hep-ph]].

  

  
 
  


  
\bibitem{Aaboud:2017mpt}
  M.~Aaboud {\it et al.} [ATLAS Collaboration],
  ``Search for long-lived charginos based on a disappearing-track signature in pp collisions at $ \sqrt{s}=13 $ TeV with the ATLAS detector,''
  JHEP {\bf 1806} (2018) 022
  [arXiv:1712.02118 [hep-ex]].
  
\bibitem{Sirunyan:2018ldc}
  A.~M.~Sirunyan {\it et al.} [CMS Collaboration],
  ``Search for disappearing tracks as a signature of new long-lived particles in proton-proton collisions at $\sqrt{s} =$ 13 TeV,''
  JHEP {\bf 1808} (2018) 016
  [arXiv:1804.07321 [hep-ex]].
  


  
  
\bibitem{ATLASprojected}
The ATLAS collaboration, ``ATLAS sensitivity to winos and higgsinos with a highly compressed mass spectrum at the HL-LHC'', ATL-PHYS-PUB-2018-031.


 
\end{thebibliography}
\end{document}